\documentclass[12pt]{iopart}
\usepackage{iopams}  
\usepackage{graphicx}
\usepackage{subfigure}
\usepackage{epstopdf}% To incorporate .eps illustrations using PDFLaTeX, etc.

\begin{document}

\title[Loading studies...]{On continuous loading of a U-magneto-optical trap (U-MOT) on atom-chip in ultra high vacuum}

\author{Vivek singh\textsuperscript{1, 2}, V. B. Tiwari\textsuperscript{1, 2} and S. R. Mishra\textsuperscript{1, 2} }

\address{\textsuperscript{1} Laser Physics Applications Section, Raja Ramanna Centre for Advanced Technology, Indore-452013, India.\\
\textsuperscript{2} Homi Bhabha National Institute, Anushaktinagar, Mumbai-400094, India}
\ead{viveksingh@rrcat.gov.in}
\vspace{10pt}
\begin{indented}
\item[]December 2019
\end{indented}

\begin{abstract}
Here, we report our studies on the continuous loading of a U-magneto-optical trap (U-MOT) on atom-chip from background rubidium (Rb) vapor generated using a dispenser source in ultra high vacuum (UHV) environment. Using the U-MOT loading curves, the partial pressure due to Rb vapor and pressure due to background gas have been estimated near the MOT cloud position. The estimated pressure due to Rb vapor increased from $\sim \; 1.4 \times 10^{-10}$ Torr to $ \sim \; 4.1 \times 10^{-9} $ Torr as Rb-dispenser current was increased from 2.8 to 3.4 A. The increase in dispenser current also resulted in decrease in loading as well as lifetime of the MOT cloud. This study is useful for magnetic trapping experiments where accurate information of pressure in chamber is important for the lifetime of the magnetic trap.

\end{abstract}

%\maketitle

\section{Introduction:}

An Atom-chip setup \cite{1, 2, 3, 4, stephen} provides a platform to manipulate cold atoms on miniaturized scale to achieve atom trapping \cite{4}, guiding \cite{5}, beam splitting \cite{6} and Bose-Einstein condensation (BEC) \cite{7}, etc for practical applications. In addition to this, an atom-chip offers advantage of tighter magnetic traps with possibility of high thermalization rates which can significantly reduce the time needed for evaporative cooling from minutes to seconds \cite{8}. This makes possible the achievement of BEC at moderate level of ultra high vacuum ($\sim \; 1 \times 10^{-10}$ Torr). Therefore, the atom-chip setups world wide are becoming popular with the single MOT loaded in an UHV chamber (pressure $\sim \; 1 \times 10^{-10}$ Torr) where MOT loading, magnetic trapping and evaporative cooling can be done at the same place. The starting point for these experiments is usually the loading of a MOT for initial cooling of atoms. A popular method to load MOT on atom-chip is loading an U-magneto-optical trap (U-MOT), which is formed by reflecting MOT laser beams on atom-chip surface and applying quadrupole magnetic field generated by bias coils and a current carrying U-shaped copper wire placed behind the atom-chip element \cite{9}. Atoms in the MOT are captured either from the background vapor in the UHV chamber or from a thermal atomic beam slowed down using a Zeeman slower device \cite{10, 11}.

For the MOT loaded from the background vapor, the loading rate is proportional to the partial vapor pressure of the atomic gas to be cooled and trapped. As a result, this pressure should be high for fast loading of MOT, but low, for a longer lifetime of the trapped atoms. To meet this conflicting requirement, one approach is to use a double-MOT setup  \cite{11, 12, 13, 14, 15},  where the first MOT is loaded in a vapor chamber at a relatively higher pressure ($\sim 10^{-8}$ Torr) and second MOT is loaded in UHV chamber ($< 10^{-10}$ Torr) by transferring cold atoms from the first MOT. An alternative approach is to load MOT in UHV chamber by using thermal atomic beam slowed down by Zeeman slower. However, both these schemes involve some complications. For example, the double-MOT setup involves difficulties of differential maintenance of the vacuum as well as transfer of cold atoms from first MOT to second MOT. The Zeeman slower loaded MOT involves the complications of design and implementation of Zeeman slower device. Thus, preparation of MOT directly in the UHV chamber by temporally controlling the atomic gas vapor pressure makes the UHV MOT loading much simplified \cite{16, 17, 18}. A fast rise in partial pressure of atomic vapor can be easily achieved by passing high current pulse in the metal dispensers  \cite{18}. However, the resistive heat generated by passing high current through dispenser needs to be managed properly to achieve fast recovery of vacuum after switching-off dispenser current \cite{18}. In addition to this, repeated high current pulses also reduce the lifetime of the dispenser. Therefore, the continuous loading of MOT in UHV chamber seems a better option for this purpose, in-spite of the continuous loading of MOT in UHV conditions being difficult due to low number density of atomic vapor. Though, a continuous MOT loading in UHV condition has been reported earlier \cite{19} in brief. However, the dependence of pressure, MOT-loading time and lifetime on dispenser current in a continuously loaded UHV-MOT needs to be investigated.\\

In this letter, we report detailed study on the continuous loading of a U-magneto-optical trap (U-MOT) of $^{87}Rb$ atoms from the Rb-atom vapor in the UHV chamber. We have investigated the effect of variation in Rb-dispenser current on number of atoms trapped in the U-MOT in UHV chamber, the rise in pressure in the UHV chamber, MOT loading time and lifetime of the MOT cloud. In addition to this, the Rb vapor pressure at the MOT position has been estimated at a given dispenser current using the MOT loading data. Our results show that the partial Rb pressure near MOT position is different from the pressure measured at sputter ion pump (SIP) position. This estimation of Rb-vapor partial pressure is important for magnetic trapping experiments to be performed in this setup, where the actual value of pressure determines the lifetime of atom cloud in the magnetic trap.

\section{Experimental setup}
The experimental setup used for this work consists of an octagonal vacuum chamber. A turbo molecular pump (TMP) (77 l/s), sputter ion pump (SIP) (300 l/s) and titanium sublimation pump (TSP) are used to achieve the UHV in chamber with base pressure $\sim \; 2.0 \times 10^{-10}$ Torr as read by SIP controller. Four independent MOT-beams, each of which is made by superposing a cooling beam and repumping beam, are used to form U-MOT as shown in figure 1. The physical size of MOT-beams are $\sim$ 18 mm in diameter. During the U-MOT operation, a current of $\sim$ 54 A is supplied to the copper U-wire in presence of bias fields ($B_{y} \sim 11 $ G and $B_{z} \sim 4$ G) to generate quadrupole field for the U-MOT operation. The details of the setup are discussed in reference \cite{21, 22}. The three Rb dispensers (two Rb/NF/4.8/17FT and one Rb/NF/3.4/12FT) are spot-welded in series on a two-pin vacuum feed-through. The feed-through with dispenser is mounted on a chamber port to place getter inside the vacuum chamber. The dispensers are located at a distance of $\sim$ 17 cm from the centre of the octagonal chamber. The Rb vapor is injected into the vacuum chamber by passing a current through this feed-through. The pressure inside the vacuum chamber was dependent on the current supplied to the dispenser. The SIP controller displayed the pressure of $ \sim \; 3.2\times10^{-10}$ Torr at the dispenser of 2.8 A and $ \sim \; 6.2\times10^{-10}$ Torr at a higher dispenser current value of 3.4 A.\\

 \begin{figure}[]
                \centering
                \includegraphics[width=8 cm]{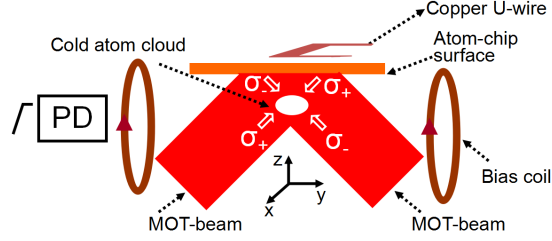}
                \caption{The schematic diagram of the experimental setup of U-magneto-optical trap (U-MOT). Two MOT-beams in reflection geometry in y-z plane are shown whereas another two MOT-beams perpendicular to this plane ($\pm$ x-direction) are not shown in this diagram. PD represents photodiode for detection of fluorescence.}
                \end{figure}

The emission of Rb vapor depends on temperature of dispenser which can be controlled by the value of current passing through dispenser. At a given dispenser current, the emission rate from the dispenser gets saturated as the thermal equilibrium is achieved at the dispenser \cite{23}. The growth of Rb atoms concentration in the experimental chamber for different values of Rb dispenser current was monitored by collecting Rb-vapor fluorescence generated by MOT-beams using a high sensitivity photodiode (PD). The growth of fluorescence signal (i.e. number density of Rb atoms in the chamber) is shown in figure 2(a) for the Rb-dispenser current of 3.4 A. The photodiode signal gets saturated in $\sim$ 150 s. It was found that saturation time is nearly same ( $\sim$ 150 s) for all values of dispenser current used during the experiment. But, the saturated photodiode signal value increased with increase in the dispenser current. It was observed that the photodiode signal of Rb fluorescence at dispenser current of 3.4 A was $\sim$ 30 times more as compared to that at dispenser current of 2.8 A. The decay of fluorescence signal starts after switching-off the Rb-dispenser current. The fluorescence decays exponentially with time constant (1/e) of $\sim$ 3.4 s, as shown in figure 2(b).\\

\section{Results and discussions}
\begin{figure}
\centering
\subfigure[]{
\resizebox*{7.0 cm}{!}{\includegraphics{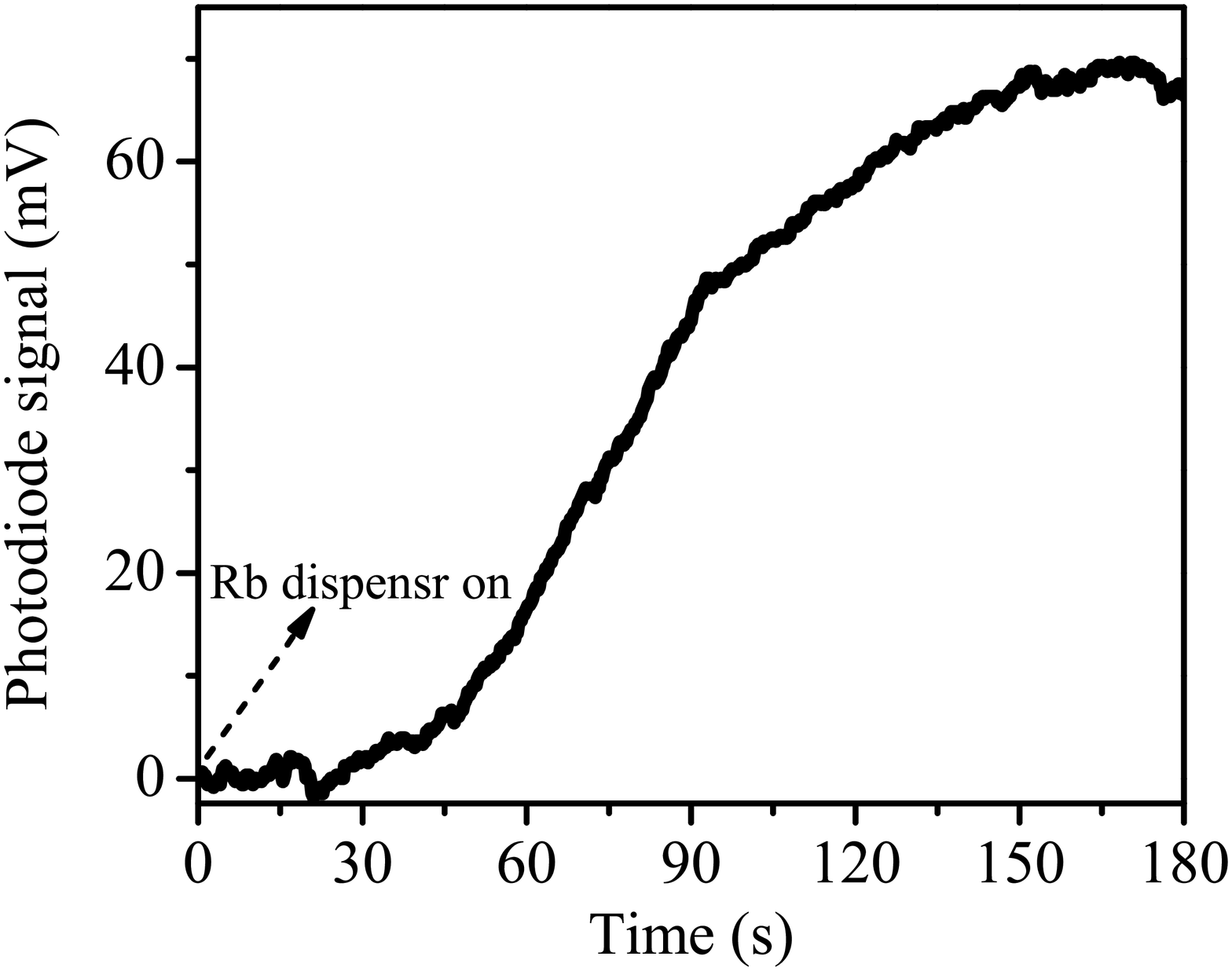}}}\hspace{0pt}
\subfigure[]{
\resizebox*{7.0 cm}{!}{\includegraphics{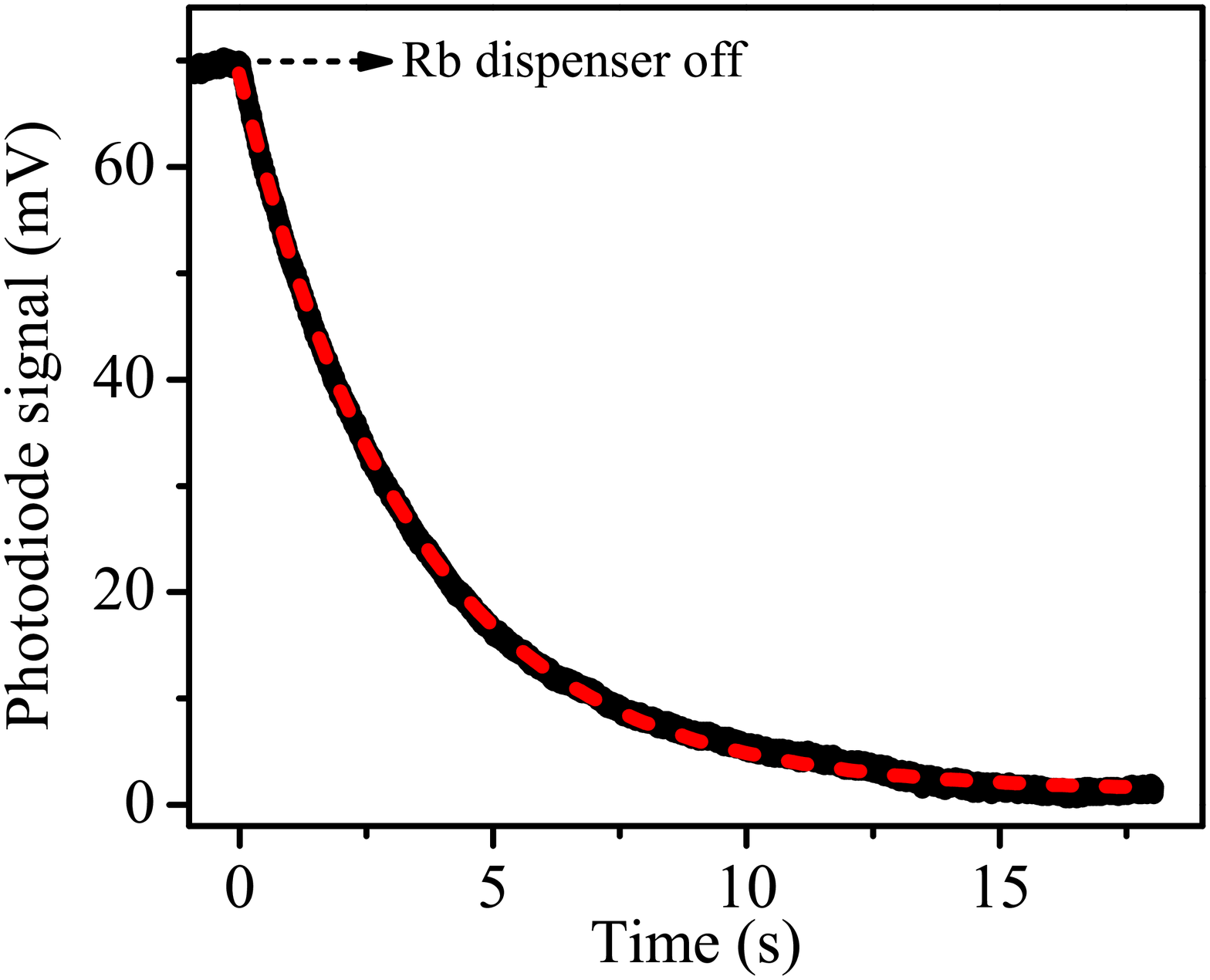}}}
\caption{(a) The photodiode signal measuring the fluorescence from background Rb-atoms in chamber for dispenser current of 3.4 A. (b) The variation in Rb fluorescence signal with time after the dispenser current of 3.4 A was switched-off. In plot (b), the experimental data are shown by continuous curve whereas fitted data are shown by dashed curve.} \label{sample-figure}
\end{figure}  
The fluorescence from the U-MOT cloud is detected using a high sensitivity photodiode (PD) as well as using a charge-coupled device (CCD) camrea. From the optical power (P) measured by this calibrated PD, we can estimate the total number of atoms N in the MOT cloud using the formula \cite{20},
\begin{eqnarray}
P = \frac{hc}{\lambda} \frac{(I/I_{sat})(\Gamma/2)}{(1+(I/I_{sat})+(2\Delta_{L}/\Gamma)^{2})} \frac{N\Omega}{4\pi},
\end{eqnarray}

where $I$ is the total intensity of all the cooling beams, $I_{sat}$ ($1.64 \; mW/cm^{2}$) is saturation intensity for the transition for circularly polarized beam, $\Delta_{L}$ is detuning of cooling laser beam and $\Omega$ is the solid angle subtended by fluorescence collecting imaging lens on atom cloud.\\

The loading of atoms in a MOT is determined by the rate equations,
\begin{eqnarray}
\frac{dN}{dt}= R-\frac{N}{\tau_{L}}
\end{eqnarray}
where $N$ is the number of atoms at any time (t) in the MOT, $R$ is the rate at which atoms are being trapped in the MOT from the background Rb vapor and $\tau_{L}$ is loading time constant. The loading time ($\tau_{L}$) of U-MOT depends upon partial Rb pressure ($P_{Rb}$) in the chamber and is given by \cite{24}, 
\begin{eqnarray}
\tau_{L}= \frac{1}{\beta P_{Rb}+\gamma_{b}}
\end{eqnarray}

where the term $\beta P_{Rb}$ represents the loss rate due to collisions with untrapped Rb atoms in the background and $\gamma_{b}$ represents collisional loss rate due to other (non-Rb) background atoms/molecules.\\
The solution of equation (2) is an exponential growth of number of cold atoms in the MOT which can be written as 
\begin{eqnarray}
N = N_{s}[1-exp(-t/\tau_{L})],
\end{eqnarray}
where
\begin{eqnarray}
N_{s} = R\tau_{L} \; =\; \alpha P_{Rb} \tau_{L}.
\end{eqnarray}
The $N_{s}$ represents the steady state number of cold atoms in the MOT and $\alpha$ is a constant representing MOT trapping cross section. After using equations (3) and (5), the relation between $N_{s}$ and $\tau_{L}$ is given by
\begin{eqnarray}
N_{s} = \frac{\alpha}{\beta}(1-\gamma_{b}\tau_{L}).
\end{eqnarray}

Since, $N_{s}$ and $\tau_{L}$ can be measured from MOT loading graph. It is obvious from equation (6) that the $N_{s}$ and $\tau_{L}$ are related, but values of $\alpha$/$\beta$ and $\gamma_{b}$ govern the variation of $N_{s}$ with $\tau_{L}$. Hence, by measuring the variation of $N_{s}$ with $\tau_{L}$ for different Rb pressure values (by varying dispenser current), it is possible to estimate the Rb pressure (from $\alpha$/$\beta$) as well as non-Rb background pressure (from $\gamma_{b}$).\\

\begin{figure}[h]
\centering
\subfigure[]{
\resizebox*{7.0 cm}{!}{\includegraphics{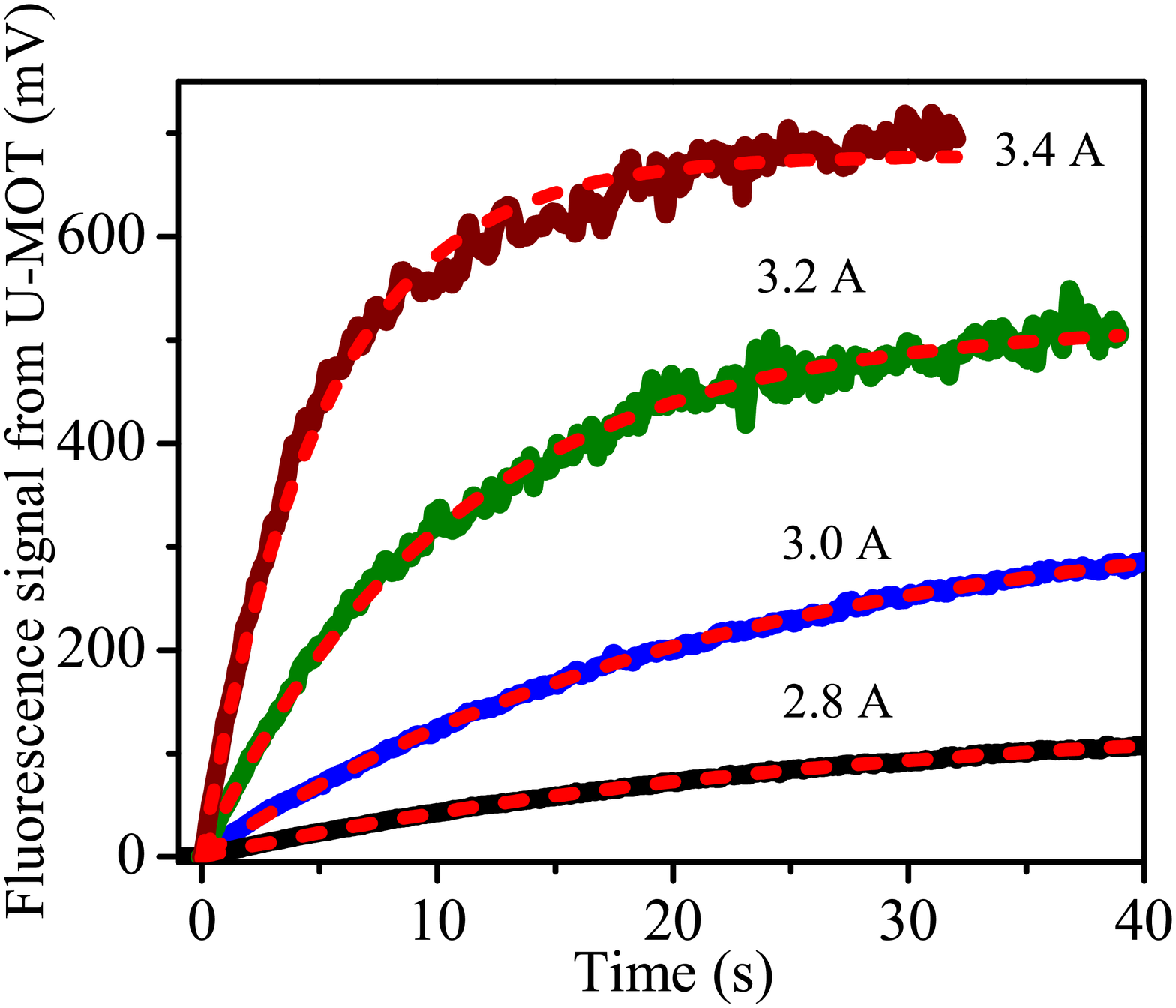}}}\hspace{0pt}
\subfigure[]{
\resizebox*{7.0 cm}{!}{\includegraphics{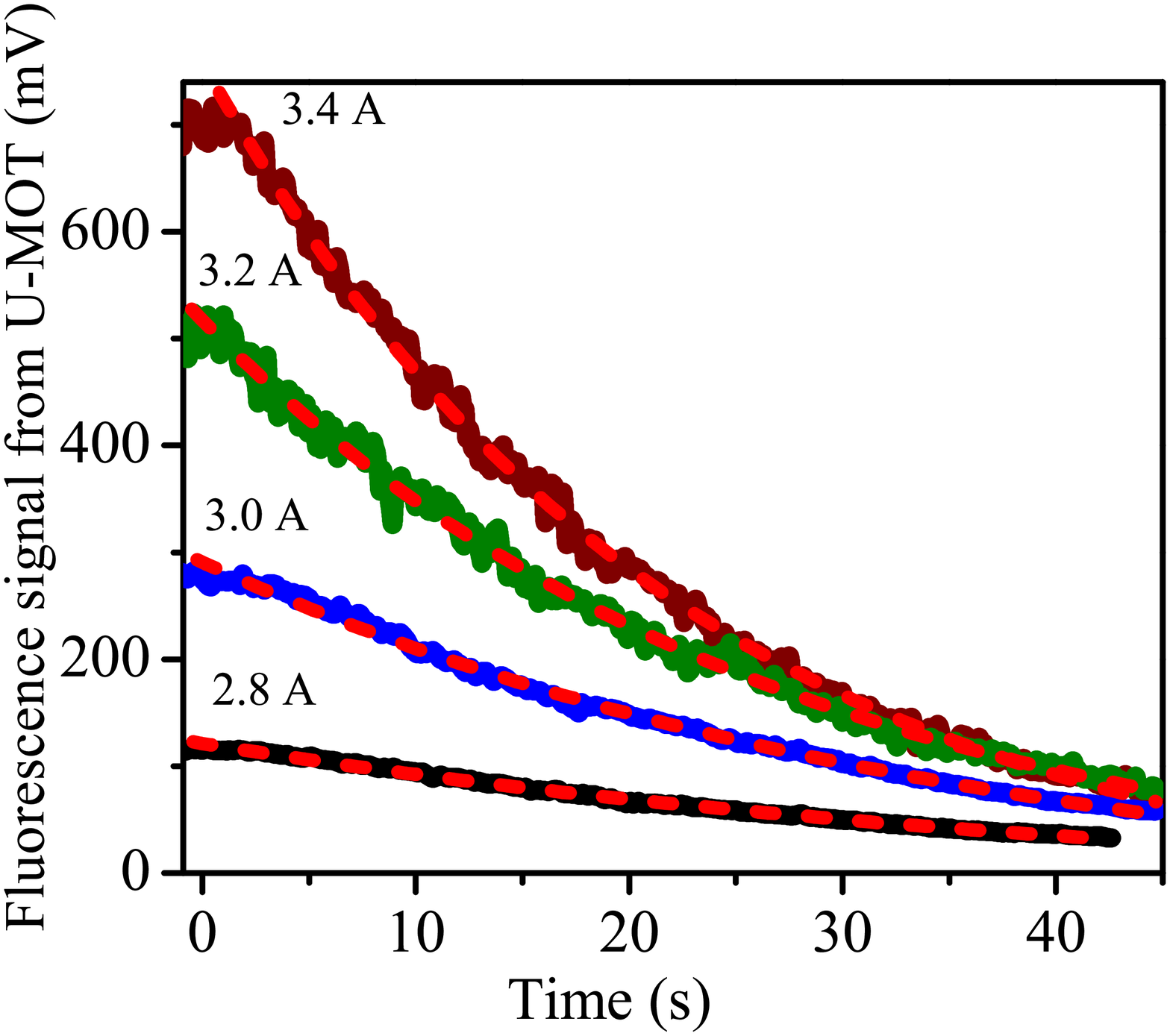}}}
\caption{(a) Loading curve of U-MOT for different values of dispenser current. The experimental data are shown by continuous curve whereas fitted data is shown by dashed curve. The trapping beams are turned on at t = 0 s and observed data is fit to equation (4). (b) The fluorescence decay curve of the U-MOT after switching-off the Rb dispenser current for different values of current. From the exponential fit of the measured data in plot (b), the MOT lifetime ($\tau_{D}$) is $\sim$ 46.8 s for Rb dispenser current of 2.8 A. The intensity of each cooling beam is $\sim$ 9 $\; mW/cm^{2}$ and detuning $\sim$ is - 12 MHz throughout the experiment.} \label{sample-figure}
\end{figure}

By measuring the fluorescence signal, the loading of U-MOT has been investigated for different values of dispenser current. The dispenser current is switched on $\sim$ 150 s before the formation of U-MOT. These values of dispenser current were chosen such that the chamber pressure remains on the order of $\sim 10^{-10}$ Torr. For MOT loading studies, the fluorescence from the U-MOT atom cloud was detected using the same photodiode (PD) as a function of time after switching on the MOT-beams. For estimation of number of atoms in the MOT, the photodiode signal was corrected after subtracting the signal corresponding to the background Rb-vapor. Figure 3(a) shows the loading curve of U-MOT i.e. rise in number of atoms in U-MOT with time, for different values of dispenser current. The observed fluorescence signal is increased as dispenser current was varied form 2.8 A to 3.4 A. The loading curves are fitted with exponential growth function to determine the loading time constant ($\tau_{L}$) for different values of  Rb dispenser current. It can be seen from the graph that $\tau_{L}$ decreases with increase in dispenser current. This is due to availability of more Rb number density in the chamber at higher dispenser current. Figure 3(b) shows the fluorescence decay curve of the U-MOT atom cloud for different values of dispenser current. This measurement was done by switching-off the dispenser current and measuring the fluorescence signal from MOT cloud as function of time, while MOT laser beams and quadrupole magnetic field were kept on. The fluorescence decay curve is fitted with exponential decay function for different values of Rb-dispenser current to determine the MOT lifetime ($\tau_{D}$). The variation in the values of loading time constant ($\tau_{L}$) and MOT lifetime ($\tau_{D}$) with dispenser current is presented in Table-1. It is evident from the table that the loading time ($\tau_{L}$)as well MOT lifetime ($\tau_{D}$) decreased as dispenser current was increased from 2.8 A to 3.4 A. \\

\begin{table}
\centering
\caption{\label{jlab1} The table of measured MOT loading time and lifetime of the MOT cloud for different values of Rb-dispenser current. All other parameters were fixed during the measurements.}
\footnotesize
\begin{tabular}{@{}lll}
\br
Dispenser current &Loading time &Lifetime  \\
(A)& ($\tau_{L}$)(s)&($\tau_{D}$)(s)  \\
\mr
2.8 A &27.72$\;\pm$ 0.07&46.8 $\;\pm$ 0.82 \\             %& $ 4.3 \times 10^{-14}$ \\
3.0 A &21.73$\;\pm$ 0.06 &36.2 $\;\pm$ 0.51\\
3.2 A & 10.53$\;\pm$ 0.05 &28.0 $\;\pm$ 0.31 \\
3.4 A & 5.12$\;\pm$ 0.03&22.2 $\;\pm$ 0.16\\

\br
\end{tabular}\\
\end{table}

The variation in MOT loading rate (R) and pressure inside vacuum chamber (as measured by SIP controller) with increase in Rb-dispenser current are shown in figure 4(a). As dispenser current was increased, the loading rate and pressure both increased. The loading rate was increased to $\sim 1.2 \times 10^{7}$ atoms/s and pressure in chamber was increased to $\sim 6.2 \times 10^{-10}$ Torr, as dispenser current was increased to 3.4 A.  Figure 4 (b) shows the variation in cold atom number with the Rb-dispenser current. As dispenser current was increased from 2.8 A to 3.4 A, the number of Rb atoms in MOT cloud increased from $\sim 1.1 \times 10^{7}$ to  $\sim 6.1 \times 10^{7}$. The enhancement in number of cold atoms is due to availability of high number density of Rb atoms inside the vacuum chamber for higher values of dispenser current. \\

\begin{figure}
\centering
\subfigure[]{
\resizebox*{7.0 cm}{!}{\includegraphics{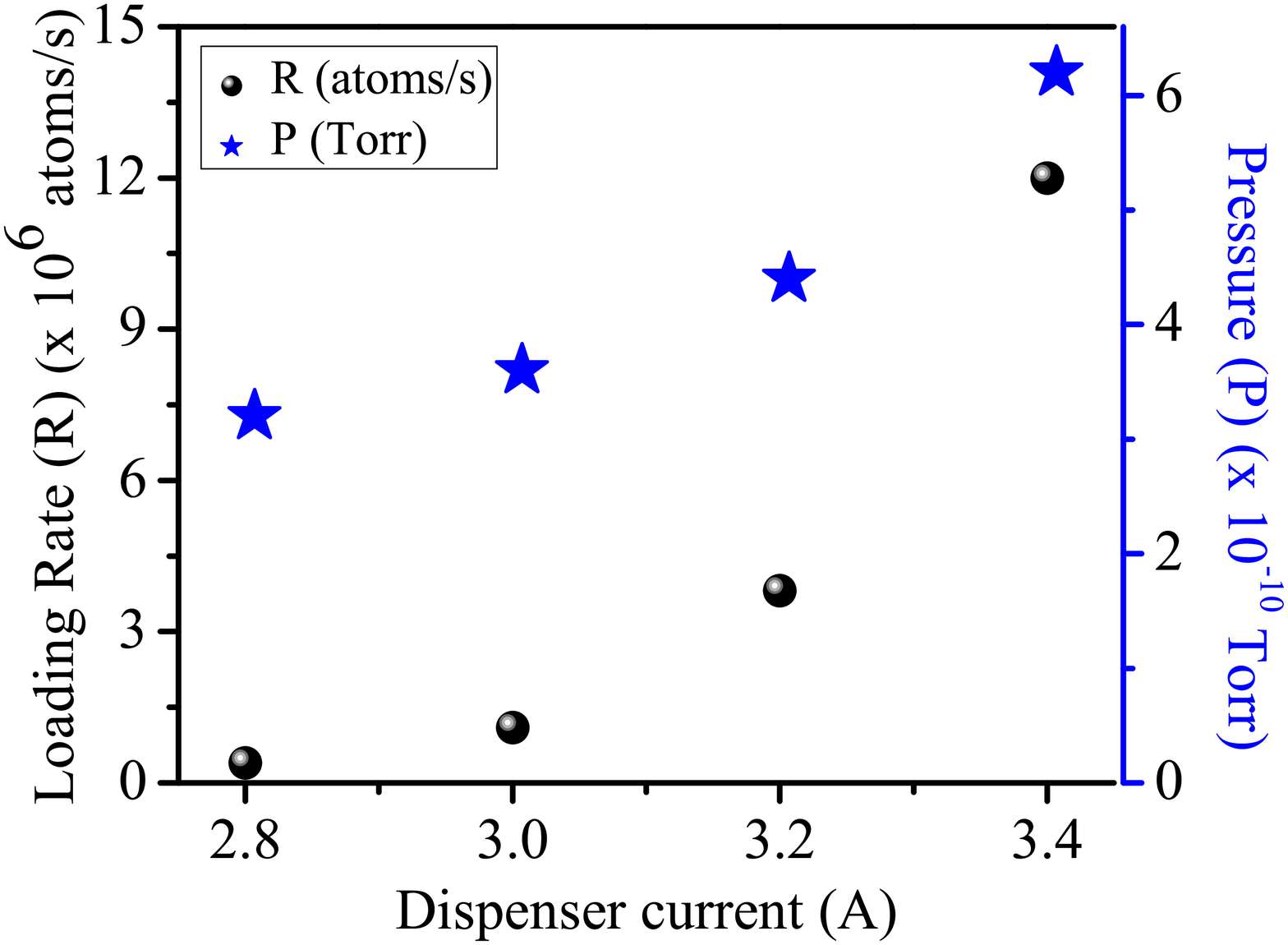}}}\hspace{0pt}
\subfigure[]{
\resizebox*{7.0 cm}{!}{\includegraphics{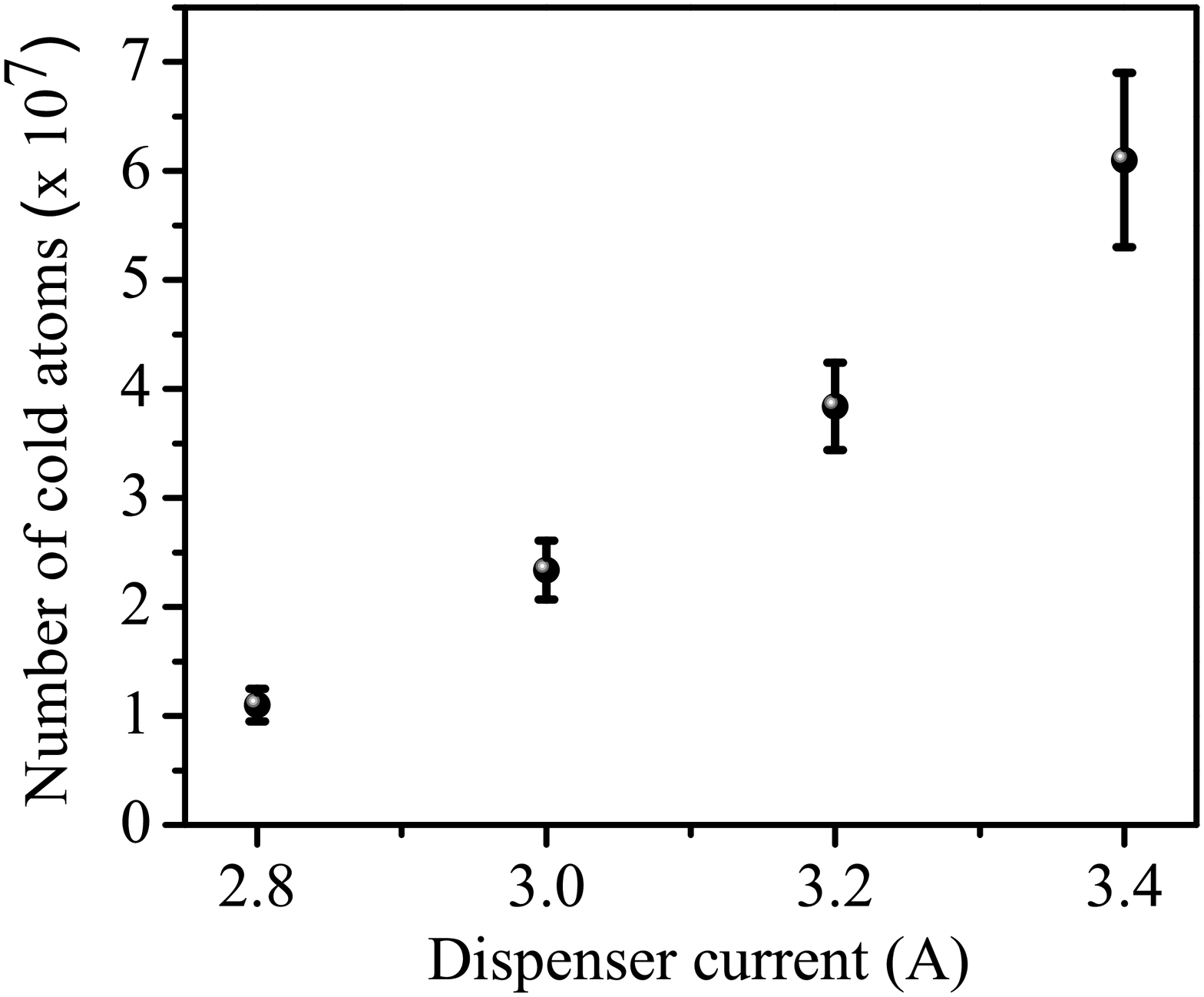}}}
\caption{(a) The variation of loading rate (R) and pressure read by SIP controller inside the chamber with dispenser current. The filled circles show loading rate and the stars symbol show pressure inside the chamber. (b) The variation in number of cold atoms in U-MOT with dispenser current.} \label{sample-figure}
\end{figure}

\begin{figure}
\centering
\subfigure[]{
\resizebox*{7.0 cm}{!}{\includegraphics{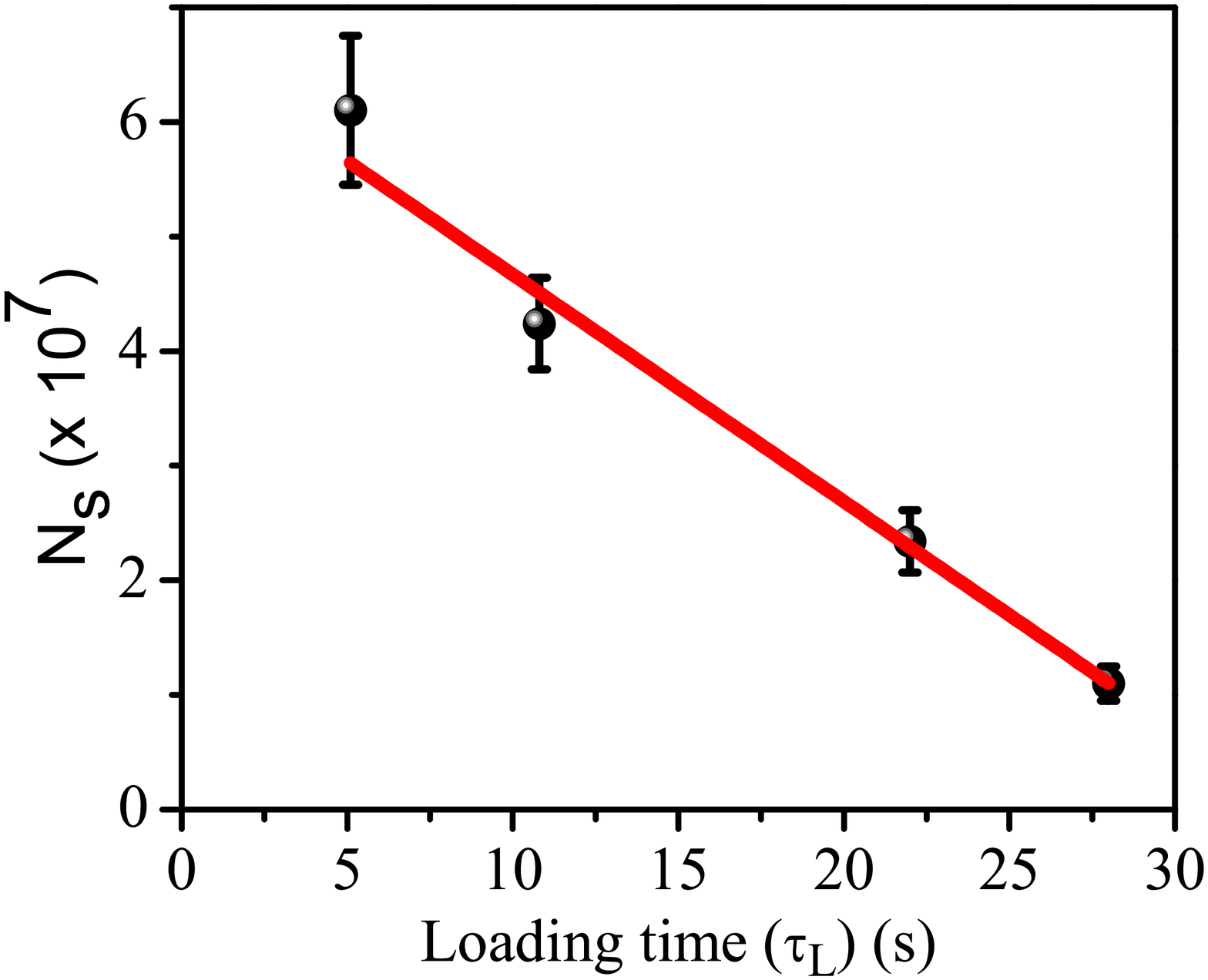}}}\hspace{0pt}
\subfigure[]{
\resizebox*{7.0 cm}{!}{\includegraphics{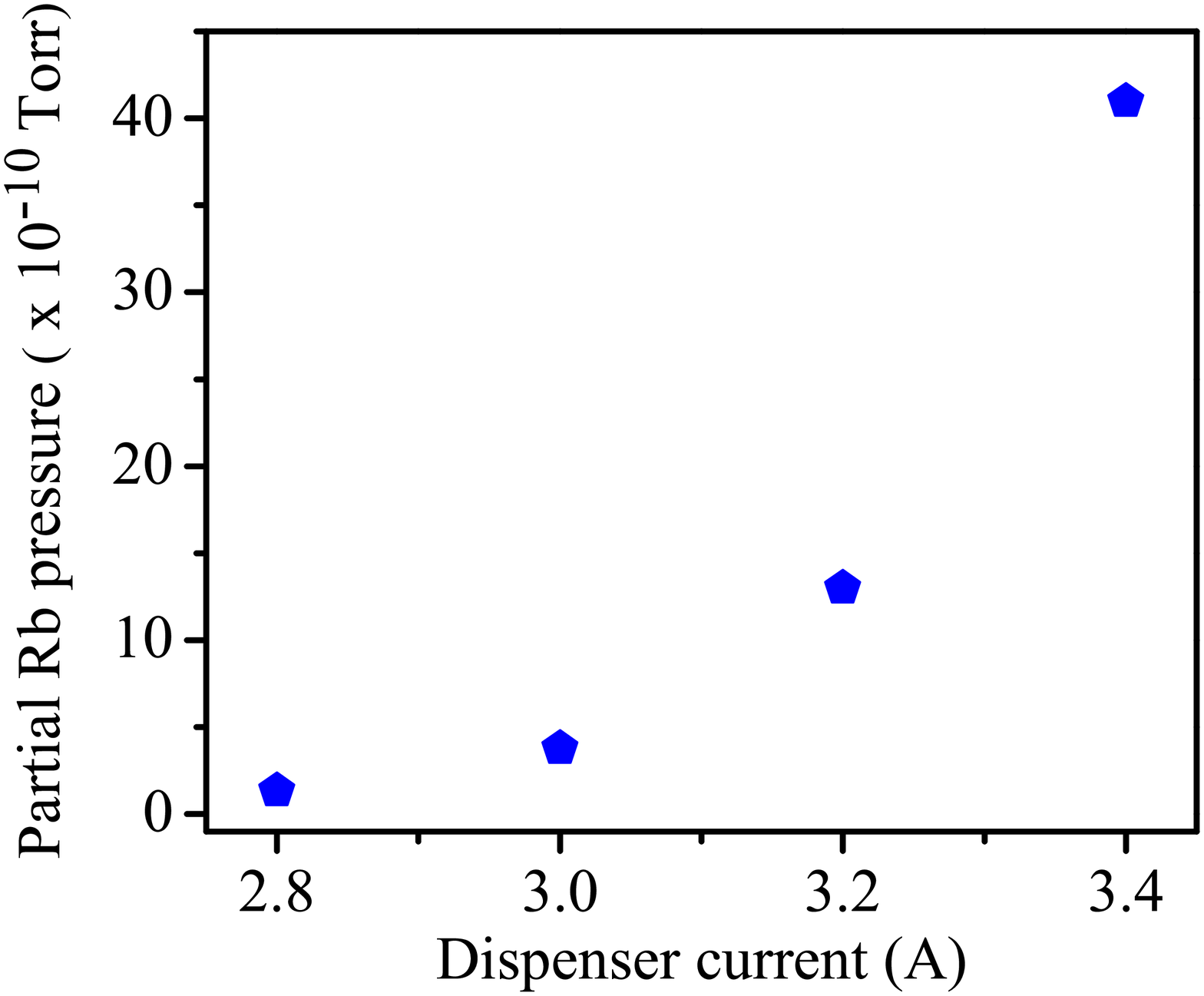}}}
\caption{(a) The variation of steady state cold atom number ($N_{s}$) with MOT loading time ($\tau_{L}$). The experimental data is fit to the equation (6). The fitting provides the values of $\frac{\alpha}{\beta} \; =\;(6.65\pm 0.3)\times 10^{7}$ and $\gamma_{b} \; =\;(2.90\pm 0.06)\times 10^{-2}$ $s^{-1}$. (b) The variation in estimated Rb-pressure with dispenser current.} \label{sample-figure}
\end{figure} 

Using the experimental data on MOT loading, we can find $N_{s}$ and $\tau_{L}$ for different dispenser currents. Figure 5(a) shows this variation in $N_{s}$ in U-MOT with loading time constant ($\tau_{L}$). Using this $N_{s} \;vs\; \tau_{L}$ plot and fitting the measured data with equation (6), we can estimate $\frac{\alpha}{\beta}$ and  $\gamma_{b}$. From the fit, we obtain $\gamma_{b} \; =\;(2.90\pm 0.06)\times 10^{-2}$ $s^{-1}$ and $\frac{\alpha}{\beta} \; =\;(6.65\pm 0.3)\times 10^{7}$. The value of $\gamma_{b}$ is related to background (non-Rb) pressure ($P$) and it is converted to pressure by using relation $\gamma_{b}/P \; = \; 4.9 \times 10^{7} \; Torr^{-1} s^{-1}$ \cite{24, 25}. The partial Rb pressure can be estimated using the value $\beta \;= \; 4.4 \times 10^{7} \; Torr^{-1} s^{-1}$ as given in literature \cite{24,25}. This gives value of $\alpha\;= (2.93\pm 0.13)\times 10^{15}\; Torr^{-1} s^{-1}$. Using these values of $\alpha$ and $\gamma_{b}$, the Rb atoms pressure as well as background gas (non-Rb atoms and molecules) pressure in the chamber can be estimated respectively. The estimated Rb pressure is $\sim \; 1.4 \times 10^{-10}$ Torr at dispenser current of 2.8 A and  $\sim \; 4.0 \times 10^{-9}$ Torr at dispenser current of 3.4 A. The estimated value of partial pressure of Rb-vapor in chamber is plotted with Rb-dispenser current in figure 5(b).

Using the value of $\gamma_{b}$, the estimated background pressure in our chamber is $\sim \; 5.9 \times 10^{-10}$ Torr. It is around three times higher than the base pressure measured by SIP controller ($\sim \; 2 \times 10^{-10}$ Torr) without any current in dispenser. This difference is due to several following possible reasons. One reason is that the method adopted above to estimate the background pressure from the $\gamma_{b}$ value is approximate with factor of 2 which depends upon the variation in trap parameters, background gas composition or trapped alkali-metal species \cite{25}. Another reason for the discrepancy between estimated pressure and SIP controller measured pressure could be the location of the SIP in the low conductance region.\\

This approach of estimating the Rb and non-Rb pressure in chamber can be helpful in optimizing dispenser current for magnetic trapping experiments, where lower pressure and sufficient number of cold atoms in the MOT are desirable parameters. For example in our U-MOT operation at the dispenser current of 3.0 A, the Rb vapor pressure $\sim \; 3.8 \times 10^{-10}$ Torr with cold atom number is $\sim \; 2.4 \times 10^{7}$. Such a parameter regime can be suitable for magnetic trapping experiments.\\

\section{Conclusion}
In conclusion, the loading of a U-magneto-optical trap (U-MOT) in continuous mode from background vapor in UHV environment is studied. It is found that the loading rate and pressure inside chamber increases with increase in the dispenser current. There is monotonic decrease in MOT lifetime and MOT loading time with increase in Rb-dispenser current. Using the MOT loading studies, the partial Rb pressure and background pressure in the chamber have been estimated. The estimated Rb pressure was increased from $\sim \; 1.4 \times 10^{-10}$ Torr to $ \; 4.0 \times 10^{-9}$ Torr with increase in dispenser current from 2.8 A to 3.4 A. These results can be helpful in optimizing dispenser current for magnetic trapping experiments where lower partial pressure of Rb atoms having sufficient cold atom number in the MOT is desirable.

\section{Acknowledgement}
We acknowledge the help provided by Shri Amit Chaudhary during the experiments. We are thankful to S. P. Ram and S. Singh for helpful discussions during this work. We are also thankful to A. Kak for fabrication of vacuum feed-throughs and R. Shukla and C. Mukherjee for fabrication of atom-chip.

\section{ORCID iDs}
Vivek Singh https://orcid.org/0000-0002-8132-8504

\end{document}